\documentclass[sigconf]{acmart}

\usepackage{soul}
\usepackage{subcaption}
\usepackage{graphicx}
\usepackage{listings}
\usepackage{xcolor}
\usepackage{balance}

\def\colorful{1}

\ifnum\colorful=1

\fi
\ifnum\colorful=0

\fi

\AtBeginDocument{%
  \providecommand\BibTeX{{%
    \normalfont B\kern-0.5em{\scshape i\kern-0.25em b}\kern-0.8em\TeX}}}

\copyrightyear{2024}
\acmYear{2024}
\setcopyright{acmlicensed}
\acmConference[SIGCSE 2024]{Proceedings of the 55th ACM Technical Symposium on Computer Science Education V. 1}{March 20--23, 2024}{Portland, OR, USA}
\acmBooktitle{Proceedings of the 55th ACM Technical Symposium on Computer Science Education V. 1 (SIGCSE 2024), March 20--23, 2024, Portland, OR, USA}
\acmPrice{15.00}
\acmDOI{10.1145/3626252.3630773}
\acmISBN{979-8-4007-0423-9/24/03}

\begin{document}

\title{AI Teaches the Art of Elegant Coding:\\ Timely, Fair, and Helpful Style Feedback in a Global Course}


\author{Juliette Woodrow}
\email{jwoodrow@stanford.edu}
\orcid{0009-0006-8097-093X}
\affiliation{%
  \institution{Stanford University}
  \state{California}
  \country{USA}
}

\author{Ali Malik}
\email{malikali@cs.stanford.edu}
\orcid{0009-0007-1201-5014}
\affiliation{%
  \institution{Stanford University}
  \state{California}
  \country{USA}
}

\author{Chris Piech}
\email{piech@cs.stanford.edu}
\orcid{0000-0001-5140-0467}
\affiliation{%
  \institution{Stanford University}
  \state{California}
  \country{USA}
}





\begin{abstract}

Teaching students how to write code that is elegant, reusable, and comprehensible is a fundamental part of CS1 education. However, providing this ``style feedback'' in a timely manner has proven difficult to scale.
In this paper, we present our experience deploying a novel, \emph{real-time} style feedback tool in Code in Place, a large-scale online CS1 course. Our tool is based on the latest breakthroughs in large-language models (LLMs) and was carefully designed to be safe and helpful for students. We used our Real-Time Style Feedback tool (RTSF) in a class with over 8,000 diverse students from across the globe and ran a randomized control trial to understand its benefits. We show that students who received style feedback in \textit{\textbf{real-time}} were five times more likely to view and engage with their feedback compared to students who received delayed feedback. Moreover, those who viewed feedback were more likely to make significant style-related edits to their code, with over 79\% of these edits directly incorporating their feedback. We also discuss the practicality and dangers of LLM-based tools for feedback, investigating the quality of the feedback generated, LLM limitations, and techniques for consistency, standardization, and safeguarding against demographic bias, all of which are crucial for a tool utilized by students.

\end{abstract}

\ccsdesc[500]{Social and professional topics~CS1}
\ccsdesc[500]{Human-centered computing~User interface programming}

\keywords{LLMs, GPT, Deployed at Scale, Real Time, Style Feedback, CS1}



\maketitle

\section{Introduction}

Establishing style principles early in computer science education teaches students to write code that is elegant, maintainable, and comprehensible. These are imperative skills for a computer scientist as they enhance software quality, minimize debugging time, and foster collaboration. 
However, providing individualized style feedback to students is a difficult task. Traditional human-grading methods are costly and slow, and do not scale to thousands of students. While automated tools exist, they are complicated to set up, can be inflexible, and often require a dataset of student solutions.

We introduce a real-time style feedback tool (RTSF) and our experience deploying it to over 8,000 learners from all around the globe. RTSF gives feedback along several key stylistic aspects such as quality of identifier names, use of constants, comments and documentation of code, and decomposition. Our tool uses cutting-edge Large Language Model (LLM) analysis to provide students with specific stylistic insights. Its main purpose is to serve as a supportive scaffold for novice programmers in this CS1 course.

We ran a randomized control trial in Code in Place, a free online course of over 8,000 students from across the globe \cite{piech2021code, codeinplace, Data2023codeinplace}. We explored the effect of timely feedback on student learning, the quality, merits, and limitations of feedback produced by an LLM, and delved into the necessary safeguards for deploying an LLM-based tool to students. 

We found that LLMs are effective at providing valuable style feedback to students, leading to noticeable improvements in their code's style. However, we also found it crucial to maintain careful control and rigorous evaluation of this process to ensure its quality and reliability. 

In summary, the main contributions of this work are:
\begin{enumerate}
    \item We created an LLM-based automated style feedback tool designed to ensure student safety and effectiveness in providing high-quality feedback.
    \item We attempt to provide the full set of feedback that human tutors would provide in a high resource classroom.
    \item We ran a large scale randomized control trial to more than 8,000 students around the world to understand how timing and LLM generated style feedback impacts learning. 
    \item We show that students who received feedback in real time were five times more likely to view it than students who received feedback a week later.
    \item We show that students who viewed style feedback had higher style scores and were more likely to make a post-functionality style edit, with 79\% of these edits directly incorporating their feedback.
    \item We investigated and found no gender bias in this particular LLM generated feedback and explored other demographic impacts of style feedback.
    \item We open-source all of our code and LLM prompts so that anyone can implement this tool in their course. 
\end{enumerate}

\begin{figure*}[t]
    \centering
    \includegraphics[width=0.8\textwidth, height={3.5cm}]{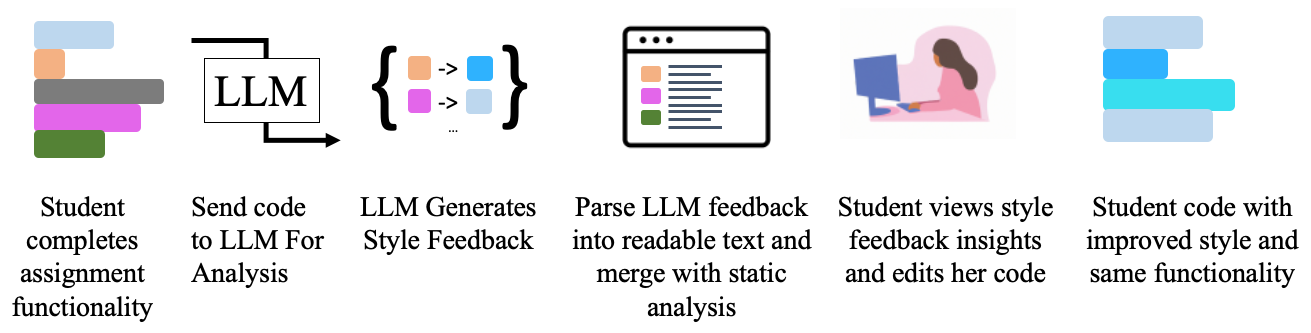}
    \caption{We employ LLM-based analysis to examine students' code and provide suggestions regarding coding style in real-time. Students can review these suggestions and make edits to improve the style of their code immediately after passing all functionality tests.}
    \label{fig:your-label}
\end{figure*}

\subsection{Related Work}

\paragraph{Timeliness of feedback}

Our tool provides instant style feedback. Prior work suggests that immediate feedback improves performance by highlighting the gap between the current output and the target goal \cite{TimingOfFeedback}. Students are more successful in applying the concepts they've learned when they receive feedback promptly, for instance, within an hour \cite{TimingOfFeedback, LitReviewFeedbackHigherEd}. In fields where immediate feedback is accessible, integrating this instantaneous formative feedback with student revision leads to improved grades \cite{goldstein2016improving, RapidFeedbackEngCourse}.

\paragraph{Automated feedback tools}
Over the past 50 years, a growing volume of CS1 education research has centered on efficient and prompt large-scale style feedback delivery. \cite{LitReviewPaper, ACMCodingGuidelines, AnalyzingStudentCode1999, StudyOfAutoStyleHelping}. One of the reasons automatic feedback is so important is that it can be delivered in real-time. Early prior work on automated style feedback relied on static analysis or rules-based tools \cite{CLinter, EarthWorm, TutoringCodeRefactoring, StaticAnalysisASTSinJava, StaticAnalysisInC, JavaAnitpatterns, PyTA, GamificationStyle, StyleAnalysisOfCPrograms, StaticAnalysisStudentJava}.

Several modern tools have been developed using NLP algorithms and clustering techniques. These techniques require datasets of pre-existing student solutions, enabling them to compare a new submission to any solution in the dataset that demonstrates better code style \cite{CharisPaper, AutoStyle, Allamanis2015SuggestingAM, OtherCharisPaper, GlassmanFoobaz}. Other miscellaneous tools focus on improving style in young learners using Scratch \cite{Pirate}, or tools that are geared towards more advanced students and software engineers \cite{RefacTutor, SoftwareQualityAssuranceTraining, HyperStyle}.

Compared to these tools, our approach does not require a pre-existing dataset of labeled solutions and instead leverages LLM-based analysis that can dynamically offer rich feedback, unrestricted by the rigidity of rule-based static analysis.

\section{Style Feedback Tool}

The style feedback tool uses LLMs and deterministic algorithms to analyze students' code and provide them with specific and individualized style feedback in real time. We present an overview of the tool in Figure 1. The source code and example LLM prompts can be found at \url{https://github.com/juliettewoodrow/realtimestylefeedbacktool.git}

The option to request style feedback is enabled only after students have successfully passed all functionality tests. Once students request style feedback, their code is sent to the LLM for analysis. The LLM response is parsed into plain text, combined with static analysis feedback, and displayed to the students. This entire process takes 5 seconds, on average. We show an example of feedback presented to students in Figure 3.

\subsection{What is Style Feedback?}

Good coding style usually denotes organized, well-documented code with meaningful variable names, and easy maintainability and debuggability. However, when teaching style in CS1, educators have differing interpretations, each holding their own perspectives on what elements are significant and what should be conveyed to students. 

We collaborated with course instructors to identify which style aspects for this tool to emphasize. Note that these can easily be updated by changing the prompts. The feedback in our tool targets the following four categories:

\begin{itemize}
    \item \textbf{Identifier Names:} Identifier names refer to the labels students assign to variables and functions in their code. Clear and meaningful identifier names enhance code readability and understanding.

    \item \textbf{Constants and Magic Numbers:} Constants are values that do not change throughout the program's execution. In contrast, magic numbers are arbitrary numerical values embedded directly in code. Using constants instead of magic numbers makes code more maintainable and flexible.

    \item \textbf{Comments:} Comments are textual explanations added within the code to provide insights, explanations, and clarifications. They offer guidance to other programmers (and to the students themselves) about the code's functionality, purpose, and tricky aspects.

    \item \textbf{Decomposition:} Decomposition refers to the process of breaking down a complex programming problem into smaller, manageable subtasks and helper functions. It aids in tackling complex problems by focusing on smaller components, which are often easier to understand and test.
\end{itemize}

\subsection{Feedback Generation}
Each of the four feedback categories has a separate process for generating feedback. Identifier Names and Comment feedback are generated by an LLM while Constants and Magic Numbers and Decomposition feedback are generated using only static analysis.

For identifier feedback, we use LLM analysis to identify any names that need improvement, suggest alternative variable names, and share a short explanation for why the alternative name is better than the one the student chose. For comments, the LLM analysis generates one positive piece of feedback relating to comments (if there were any comments in the student program) and two additional places they could add comments (if necessary). 

The feedback for static analysis of constants and magic numbers displays a list of defined constants that are never used, numbers that should be converted into constants, values defined as constants but used as variables, and constants that should have been written in uppercase. The decomposition feedback suggests students decompose functions that have more than 15 lines of code. Additionally, for any functions with more than four consecutive lines of code in common, the static analysis suggests they decompose that repeated code into a helper function.

We prevented students from seeking feedback on the same problem within 10 minutes, both to encourage critical thinking about the feedback and to avert model overuse.

\begin{figure}[t]
    \centering
    \begin{lstlisting}[label={fig:python}]
def main():
    # ask the user for a weight on earth
    weight = input("Enter a weight on earth:")
    weight_str = float(weight)

    z = weight_str * 0.378 
    s = str(z)

    print("The equivalent weight on Mars is " + s)
    \end{lstlisting}
    \caption{Example Student Program Before Style Feedback}
    \label{fig:enter-label}
\end{figure}

\begin{figure}[t]
    \centering
    \includegraphics[width=0.5\textwidth]{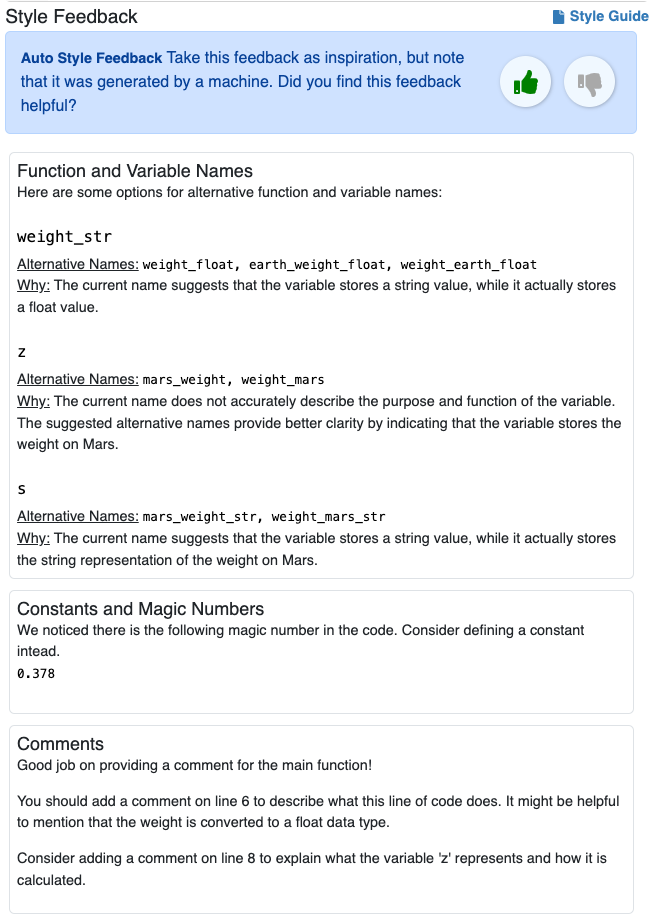}
    \caption{Example Style Feedback Presentation for the program in Figure 2.}
    \label{fig:your-label}
\end{figure}

\subsection{Prompt Design}
Our first important insight in prompt engineering was to instruct the LLM to produce output as a JSON conforming to the schema detailed in the prompt. This insight guaranteed uniform feedback by standardizing content generated and mitigating student exposure to unconstrained LLM content. It also allowed for systematic parsing and validation of generated feedback before sharing it with students.

Each type of feedback has its own prompt outlining a unique schema with specific keys. Some of these keys were designed to be used in the output shown to students and some were uniquely designed to enhance the model's output quality. For instance, we directed the LLM to calculate a score for every variable name and include that in the JSON output. We did not show the score to students, but requiring a score significantly improved the LLM's ability to distinguish between robust and weak variable names. We also directed the LLM to determine if the variable name misleads the type of information that it stores. Without including this in the schema, the LLM would not recognize instances when the name contradicted the type of value stored. 

Our second key insight was preprocessing each student's code to send additional information to the LLM. Initially, when only the code text and prompt were provided, issues arose with the generated output, including incomplete consideration of identifier names and inaccurate line number information. However, incorporating additional data notably improved the quality of the generated output. For example, for identifier feedback we sent a dictionary mapping each variable to its assigned values. For comment feedback we sent a dictionary mapping line numbers to existing comments.

\subsection{Model, Cost, Server Time Delays}

We used OpenAI's GPT-3.5-turbo model for all LLM analysis \cite{OpenAI2022GPT35Turbo}. This model was fast, cost effective, and with sufficient prompt engineering provided high quality feedback.

The total cost to use this model throughout our deployment was \$90. This was less than half a cent per student who used our tool. Model cost depends on the number of tokens in the input and in the generated output, which varies by each request. Throughout the five weeks, we had just under 56,000 requests sent to the LLM. 

Each request to the LLM took on average 5 seconds to complete.  Variations in internet access speeds and the model usage can occasionally result in longer processing times, but we did not experience any noticeable delays throughout the course. 

\section{Experiment Design}
We deployed our tool in Code in Place, a large-scale, open-access, on-line CS1 course. The tool was integrated into the students' IDE \cite{tjsIde}. Leveraging the tool's near-instant feedback generation, we investigated if feedback timing impacted student engagement. In our randomized control trial (RCT), we divided students into groups: Group \textbf{\textit{Real Time With Nudge}}, with real-time feedback and prompting upon passing functionality tests; Group \textbf{\textit{Real Time}}, with real-time feedback but no prompts; and Group \textbf{\textit{Delay}}, with the previous week's feedback available every Monday.

At the start of the course, groups were randomly assigned, with 45\% of students placed in group A, another 45\% in group B, and the remaining 10\% in group C. These group assignments did not change throughout the course. See Table 1 for student group numbers. 

Our tool was available for 5 weeks of the 6 week course. The course had a total of 15 assignments and our tool was available for 12. The course did not have any grades and the only mechanism for offering style feedback was through our tool.  
\begin{table}[t]
\centering
\small
\begin{tabular}{|p{3cm}|p{1cm}|p{1cm}|p{2cm}|}
\hline
Group & Delay & Real Time &  Real Time With Nudge \\
\hline
Number of Students & 850 & 3265 & 4195 \\
\hline
Group Use Percentage & 7.27\% & 6.56\% & 44.69\% \\
\hline
\end{tabular}
\caption{Experiment Group Details}
\label{my-label}
\end{table}

\section{Results}

\subsection{Timeliness of Feedback}
We found that if the feedback is not timely, almost nobody views it. We also found that nudging people to view feedback in real time has the highest chance of getting students to look at their feedback. We show these results in Figure 4.

\begin{figure}[t]
    \centering
    \includegraphics[width=0.4\textwidth]{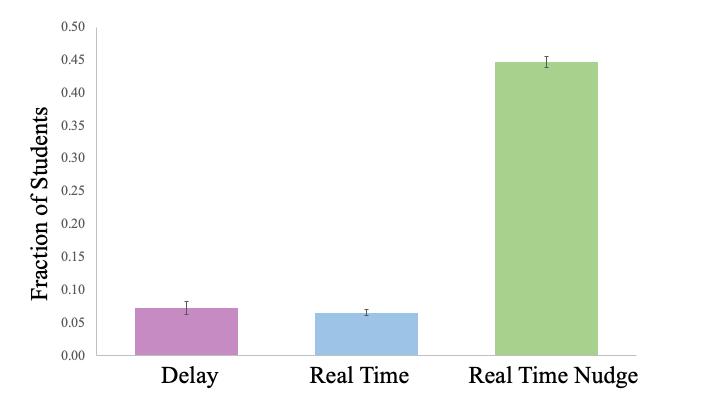}
    \caption{Students who were nudged and had timely feedback were much more likely to to view style-feedback compared to students in the other experiment groups.}
    \label{fig:your-label}
\end{figure}

The real time group was much more likely to provide a rating for the feedback. About 30\% of students in that group provided a rating with 97\% rating it helpful. The delay group almost never gave a rating (<3\%). Although definitive conclusions cannot be drawn due to the delay group's negligible ratings, it does suggest that real-time feedback viewers are more likely to value the feedback.

\subsection{Impact on Student Style}

We know that students who receive feedback in real time look at the feedback. Now we investigate how they engage with and incorporate this feedback after viewing.

Instructors of the course chose three assignments representative of the overall course work, and they hand graded a style score for a sample of students. While grading, they had no knowledge of the students' experiment groups or whether they viewed style feedback for any of the assignments. The first assignment was given and completed by students before the tool was released. The second and third assignments were spread out throughout the course after the tool was released to students.  In this section, we only evaluate students who had real-time style access and were prompted to view it, as low interaction rates among other groups prevented extensive comparative analysis between them.

On the latter two assignments, we noticed an increase in style scores for students who viewed their feedback compared to non-viewers. Figure 5 shows the Z-score of style distributions between these two groups per assignment. A slight difference was noted pre-tool release among future feedback viewers. However, these viewers exceeded the original distribution in the next two assignments, suggesting the tool's feedback contributed to their style score improvement.

\begin{figure}[t]
    \centering
    \includegraphics[width=0.5\textwidth]{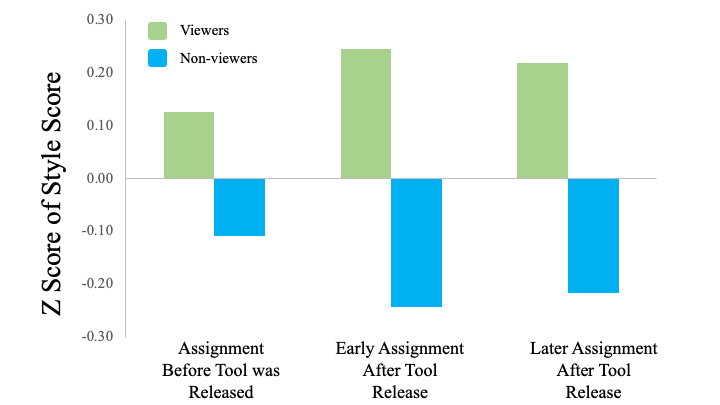}
    \caption{Feedback viewers improved their style scores more so than non feedback viewers. Z-scores of style score on three assignments throughout the course.}
    \label{fig:your-label}
\end{figure}

We examined student engagement with style feedback on the latter of the three assignments. We found that feedback viewers were more likely to make edits to their code after achieving full functionality than non feedback viewers. Figure 6 demonstrates these findings, mapping the amount of significant edits from the first code solution that clears all test cases across various time periods for both feedback viewers and non-viewers.
\begin{figure}[t]
    \centering
    \includegraphics[width=0.4\textwidth]{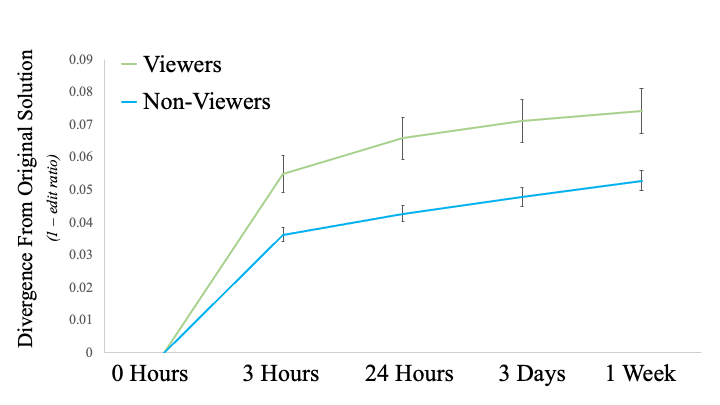}
    \caption{Students who viewed feedback made more post-functionality code edits compared to those who did not.}
    \label{fig:your-label}
\end{figure}

It could be that students who are more engaged are likely to revise their code after ensuring functionality, and these are the same students who tend to view style feedback. To control for this, we considered only the students who edited their code after they passed all functionality test cases and explored the nature of their edits. We found that of the students who made edits to their code after passing all functionality test cases, those who viewed style feedback were much more likely to make a style based edit or a combined style and functionality edit than those who did not view style feedback, shown in Figure 7. 

\begin{figure}[t]
    \centering
    \includegraphics[width=0.4\textwidth]{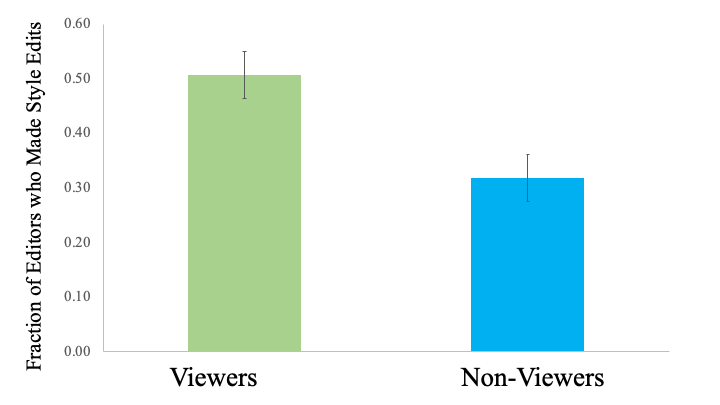}

    \caption{Considering only students who made an edit post full-functionality, a larger percentage of those who viewed style feedback made a style-based edit compared to those who did not view style feedback.}
    \label{fig:your-label}
\end{figure}

Now considering only the students who viewed style feedback and made a style based edit or a combined style and functionality edit, we found that 79\% of those students directly incorporated the style insights shown in their feedback. 

\section{Practicality of LLMs to Provide feedback}

In this section, we examine cases of both valuable and unhelpful LLM-generated feedback to assess the practical use of LLMs in generating feedback.

\subsection{Positive LLM Feedback Examples} 

The LLM-generated feedback strengths included providing useful explanations for name improvement suggestions, accurately identifying misleading variable names, and suggesting thoughtful, descriptive comments.

In Figure 2 we show an example student program before requesting any style feedback. Figure 3 shows the LLM generated feedback for that code. The LLM feedback noted that the weight\_str variable misrepresented the type of value that variable stored and recommended alternative names that were more accurate. It also suggested names for z and s that are more accurate and descriptive given the context of the program.  

The LLM also provided helpful comment feedback for this program with one positive aspect and two suggestions where the student could place a comment. The first suggestion was, "Consider adding a comment on line 4 to explain the purpose of converting the user input to a float." The second suggestion for comments was, "On line 7, you could add a comment to describe the calculation being performed and why the value 0.378 is used. This would provide more clarity to the reader." The positive feedback was, "Great job adding a comment on line 3 to describe what that code does." These examples are indicative of the quality of most of the feedback generated in our deployment.

\subsection{Limitations of LLM Generated Feedback}
During the instances where the feedback was unhelpful, potential issues included: not highlighting all identifier names that needed improvement, generating feedback about comments that did not exist, and inconsistent variable name suggestions. 

In certain cases, the model inaccurately gave some variable names high scores even though they needed improvement, wrongly implying they were good choices. For example, with the program in Figure 2 as input, there may be instances where the model incorrectly assigns high scores to 'z' or 's', suggesting they are descriptive variable names when they are not suited for the given context.

An example illustrating hallucinated comments is when the LLM praised a student's program, which lacked any comments, with "Great job adding comments to the program!" Or it would suggest adding a comment that was not particularly helpful or necessary. 

The alternative variable names suggested by the LLM were inconsistent. The generated feedback sometimes offered more accurate and descriptive names and sometimes the names were no better than the existing one chosen by the student, or were overly descriptive.

\begin{figure*}[t]
\begin{subfigure}{0.31\textwidth}
    \centering
    \includegraphics[width=\linewidth]{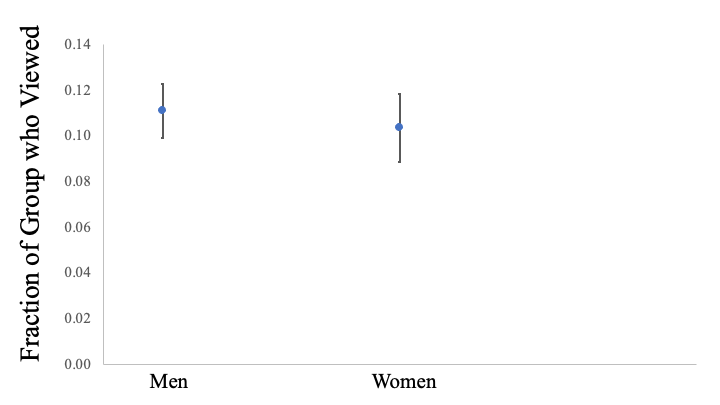}
    \caption{The usage of the tool among men and women is consistent.}
    \label{fig:your-label}
\end{subfigure}
\hfill
\begin{subfigure}{0.31\textwidth}
    \centering
    \includegraphics[width=\linewidth]{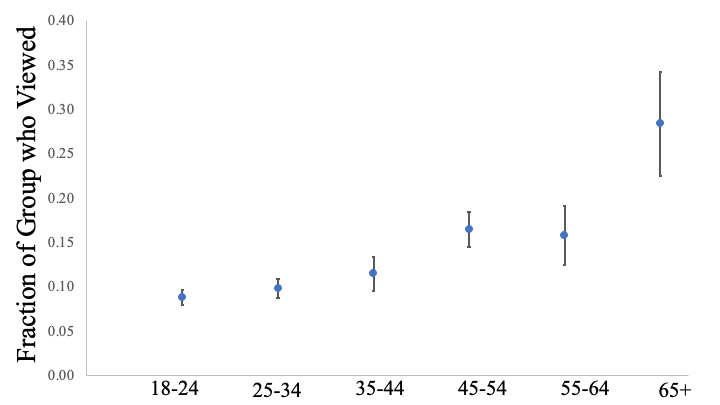}
    \caption{The fraction of students in an age group who use the tool trends upwards with age.}
    \label{fig:your-label}
\end{subfigure}
\hfill
\begin{subfigure}{0.31\textwidth}
    \centering
    \includegraphics[width=\linewidth]{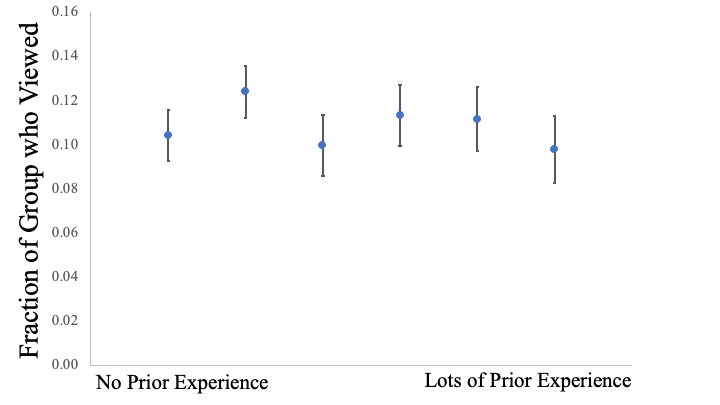}
    \caption{Usage of our tool is consistent across students with varying levels of prior experience.}
    \label{fig:your-label}
\end{subfigure}
\caption{Fraction of Students in Different Demographic Groups Who Viewed Style Feedback}
\end{figure*}

\section{Safety and Bias}

In this section, we first outline our strategies for ensuring fair and safe LLM-generated feedback, then we analyze feedback utilization among different demographic groups.
\subsection{Ensuring Fairness and Safety in LLM Generated Feedback}
A valid and significant concern when using LLMs to generate text for students is the potential bias in the output, which could negatively affect specific groups. We were highly conscious of this issue, as we aimed to avoid perpetuating any biases that might hinder certain groups from feeling welcome in programming. To address this concern, we put in place several safeguards:
\begin{itemize}
    \item We shared only the text of each student's code, devoid of any information about the student or their course progress.
    \item Students were limited to one round of LLM interaction at a time. 
    \item Students' interactions with the LLM were solely via our comprehensive parsing and validation system.
    \item To avoid unsupervised LLM generated plain text, the LLM was set to deliver responses in JSON form.
\end{itemize}

We evaluated the safety of our generated feedback through rigorous checks of thousands of responses, finding no instances of inappropriate or demographically biased content. Using a sentiment analysis classifier on a subset of feedback provided to men and women, we found that the model did not provide differential feedback based on gender \cite{huggingfacemodel}. A manual examination of sampled feedback given to male, female, and non-binary students confirmed no discrepancies. 

\subsection{Demographic analysis}

\paragraph{Gender Distribution and Tool Usage}

Our study demonstrated an equitable usage of our tool by both men and women, with 11\% of men and 10.4\% of women utilizing it, indicating its universal appeal and accessibility, as detailed in Figure 8. However, usage was less among non-binary students at only 4.1\%. A comprehensive analysis of the tool and feedback generated showed no differences in feedback across genders, but the lower usage by non-binary students calls for further investigation.

\paragraph{Age and Tool Adoption}
A notable trend emerged in the analysis of tool usage across age groups, displayed in Figure 9. The fraction of students using the tool displayed an upward trajectory with increasing age. This observation might represent students' changing needs and preferences throughout their education. It underscores the need to customize tool interfaces and functions for different age groups.

\paragraph{Prior Programming Experience and Tool Engagement}

We investigated tool usage across varying prior programming experience levels, revealing consistent engagement between 9\% to 13\% for all groups. This indicates our tool's versatility, benefiting beginners and experienced programmers alike. Further research is needed to determine if there is an upper limit on experience level beyond which the tool's ability to provide effective style feedback decreases.

\section{Limitations of the Tool}

Section 5.2 discussed the constraints of LLM-generated feedback. This section focuses on key factors to consider in the broader application of the tool. 

The feedback quality from the LLM is inherently connected to its training data. In the case of the model we used, the training data is non-public and non-adjustable. This limits flexibility, as the tool cannot adapt to diverse style preferences. Feedback from this tool mainly serves as formative rather than summative. As described in Section 5, the LLM generated output varies, leading to potential application restrictions outside of formative feedback provision.

The tool's static analysis may overlook some errors due to its strict structure, indicating a need for more context-sensitive strategies to ensure thorough feedback. Additionally, although the tool offered feedback on four main style categories, it cannot generate high-quality feedback in other important areas such as indentation and major code restructuring.

\section{Further Research}
Our tool facilitates further exploration into LLM feedback generation, its effect on various demographics, fairness, bias, and the creation of LLM-based resources for learners. Future research should conduct a more thorough investigation into the potential biases of LLM-based tools than was accomplished in this report. It is also crucial to examine in greater depth how the usage of these tools, as well as the nature of the responses they produce, vary among different demographic groups.

Future research should explore open-source models. The investigation of open-source LLM models can promote transparency, accessibility, and collaborative advancements. This would allow instructors to tailor these models to their specific style guidelines, diminishing costs and reliance on external platforms.

\section{Conclusion}

In this experience report, we investigate the influence of timely and practical LLM-generated feedback on a large group of students worldwide. We found timeliness of feedback to be crucial. Students who received real-time feedback were five times more likely to view it than those with delayed feedback. Furthermore, students viewing the LLM-generated feedback achieved higher style scores and were more likely to make style-based edits. We also analyzed and summarized the advantages and limitations of using a cutting-edge LLM to generate feedback. Our demographic analysis demonstrates the tool's versatility, highlighting its wide applicability across various demographics. We invite you to use and build upon this novel tool to provide real-time style feedback in your CS1 courses.

\begin{acks}
We would like to sincerely thank the Carina Foundation and the Code in Place team for making this research possible.
\end{acks}

\clearpage
\bibliographystyle{ACM-Reference-Format}
\balance
\bibliography{ref}


\end{document}